# Towards exhaustive branch coverage with PathCrawler

Nicky Williams
CEA, LIST,
Université Paris-Saclay,
F-91120 Palaiseau, France
nicky.williams@cea.fr

*Abstract*— Branch coverage of source code is a very widely used test criterion. Moreover, branch coverage is a similar problem to line coverage, MC/DC and the coverage of assertion violations, certain runtime errors and various other types of test objective. Indeed, establishing that a large number of test objectives are unreachable, or conversely, providing the test inputs which reach them, is at the heart of many verification tasks. However, automatic test generation for exhaustive branch coverage remains an elusive goal: many modern tools obtain high coverage scores without being able to provide an explanation for why some branches are not covered, such as a demonstration that they are unreachable. Concolic test generation offers the promise of exhaustive coverage but covers paths more efficiently than branches. In this paper, I explain why, and propose different strategies to improve its performance on exhaustive branch coverage. A comparison of these strategies on examples of real code shows promising results.

## I. INTRODUCTION

### A. The importance of exhaustive branch coverage

*Structural coverage criteria* are based on the simple observation that if the part of the program containing a bug is not executed (a.k.a. *covered*) by any test case then the bug cannot be detected. Various structural coverage criteria defining *test objectives* in source code, binary code and models or specifications have been proposed. The structural code coverage criteria mostly widely used in industry seem to be line coverage, branch coverage and MC/DC. These are often imposed by certification norms. Moreover these norms often impose 100% coverage of coverable test objectives and, by extension, some justification of why certain test objectives cannot be covered. We call this *exhaustive coverage* and define it as the generation, for all test objectives, of either (a) a test case input covering the objective or (b) an explanation of the lack of a test case. We insist on the difference between exhaustive coverage and *bug finding* i.e. the speedy coverage of many test objectives with no guarantee of completion.

Automatic test input generation techniques for exhaustive coverage must keep track of which objectives have already been covered in order to know when to stop test generation and to try to avoid generating numerous test cases which cover the same objectives while failing to cover others. Moreover, in order to provide an explanation of failure to cover a particular test objective, test generation should not stop until all possible attempts to cover the objective have been made and should then report on the result of these attempts. Let us suppose that the uncovered objective is a branch in the source code. If there seem to be several partial paths through the source code leading to the branch, then all these paths should be considered. If the infeasibility of all paths to a branch can be automatically demonstrated, then this is the explanation of failure to cover the branch. If infeasibility of one or more paths cannot be demonstrated automatically, for example, because a constraint solver times out, then the explanation for these paths is the formula whose satisfiability the solver was trying to prove or disprove.

We focus on the exhaustive coverage of branches because the problems that it poses can be generalised to several other problems. Indeed, if we consider that each branch leads to a, possibly empty, sequential block of lines then the only difference between line coverage and branch coverage is the coverage of "empty" branches. MC/DC imposes the further requirement to cover certain pairs of combinations of atomic branches, which essentially involves additional book-keeping. Moreover, the coverage of other test objectives which can be defined in the source code (e.g. assertion violations), or even as pseudo-branches (e.g. run-time errors such as division by zero) or source code annotations [1], is also similar to branch coverage. Techniques for branch coverage can be extended to these test criteria and indeed already are in some automatic test generation tools such as Pex/Intellitest [2], KLEE [3] and PathCrawler [4]. Other verification tasks apart from testing, often framed in terms of *reachability*, also pose problems similar to exhaustive branch coverage. In these, unwanted program states are defined and the problem is to demonstrate whether they can occur. If they can, then the user often needs a counter-example, i.e. test inputs, to help with debugging.

### B. Techniques for branch coverage

Static analysis based on abstract interpretation can be an efficient technique for detecting unreachable branches but it cannot generate test input values and because of the over-approximation which is inherent in this approach, it cannot guarantee detection of all unreachable branches. It can be used in exhaustive branch coverage as a prelude to automatic test generation in order to reduce the number of test objectives [5].

Most test generation techniques currently used for branch coverage fall into four categories

1. Search-based testing is based on meta-algorithms, such as genetic algorithms, which can only be used for bug-finding because they cannot guarantee any results of test generation or provide any justification for uncovered objectives.

2. Fuzz testing can rapidly cover branches on large code bases. It often incorporates similar techniques to search-based testing or uses some symbolic execution. It can only be used for bug-finding.

3. The recent DiffBlue Cover tool for Java [6] uses model-checking. This technique could, in theory, be used for exhaustive branch coverage but DiffBlue Cover does guarantee that all reachable branches will be covered and

This work was partially supported by ANR grant ANR-18-CE25-0015-01.

does not seem to provide any justification for uncovered branches.

4. The Pex/Intellitest, KLEE and PathCrawler tools are all based on symbolic execution. These techniques can be used for exhaustive branch coverage but only if there is no "concretisation" of branch constraints and if a systematic test generation strategy, rather than one based on heuristics or randomisation, is used. To our knowledge, none of the strategies proposed for KLEE offer exhaustive branch coverage and only PathCrawler can provide an explanation for uncovered branches.

## C. Concolic generation and hopeful flipping

This paper considers exhaustive branch coverage in PathCrawler, which uses a concolic test generation method. In this method, the first test case is generated by an arbitrary choice of test inputs which satisfy the precondition (which is supplied by the user and encodes the test context). For this and each subsequently generated test case, symbolic execution is used to translate the conditions of the successive branches in the path, $p$, taken by the test case, into constraints over the input variable values. The resulting conjunction of constraints, $c_0, c_1,...$ is the path predicate, *pred(p)*, which defines the input values of all test cases which would cover $p$. In practice, the source code must be normalised, in order to unroll loops, separate out side-effects, decompose complex conditions, etc. Moreover, variables used when accessing array elements or dereferencing pointers give rise to additional alias constraints. System calls must also be stubbed. But we can consider here, without loss of generality, that each of the constraints $c_0, c_1,...$ in *pred(p)* represents the condition of one of the branches $b_0, b_1,...$ in $p$. In order to generate test inputs that will cover a different path, one of the constraints, $c_i$, must be negated (a.k.a. *flipping* the branch $b_i$). The result, *pred(flip(p,i))*, is the conjunction of the prefix of *pred(p)* up to $c_{i-1}$ and the negation, $-c_i$, of $c_i$. It is the predicate of the path prefix, *flip(p,i)*, formed by appending the opposite branch, $-b_i$, of $b_i$ to the prefix up to $b_{i-1}$ of $p$. *pred(flip(p,i))* is submitted to a constraint solver. A solution to this formula gives the input values of a new test case which will cover one of the feasible paths with prefix *flip(p,i)*.

If the constraint solver finds that *pred(flip(p,i))* is unsatisfiable then we have the demonstration that *flip(p,i)* is an infeasible path prefix. If the constraint solver times out, then *pred(flip(p,i))* can be used, if necessary, in an explanation of non-coverage of either $-b_i$ or some branch which could, in theory, be reached from $-b_i$.

Constraint solving is NP-hard and may run until timeout so in order to limit worst-case exhaustive test generation time, we should limit the number of solver calls and certainly avoid calling the solver several times to solve the same problem. Another reason for limiting constraint solving is to limit the number of generated tests and hence the time to treat each test.

Classic concolic test generation reduces the number of solver calls for path coverage but may be less efficient for branch coverage. Indeed, the concolic method interleaves generation of new tests (each time a branch is flipped) and exploration of the paths covered by the previously generated tests. Covered paths are feasible and so are all their prefixes so by flipping a single branch in a feasible prefix, the concolic method limits solver calls by ensuring detection of the shortest prefixes of all the infeasible paths. The unique feature of the concolic method is that we do not know which path will be covered by the solution recovered from the constraint solver. In the case of path coverage, this does not matter because all feasible paths must be covered in any case and the concolic method ensures that constraint solving is only performed once for each node in the tree of feasible execution paths (*FEP tree*). However, we do not usually need to cover all feasible paths in order to cover all reachable branches. This is because most branches occur in several paths (although exhaustively trying all paths to an unreachable branch can sometimes necessitate full path coverage).

In concolic test generation for branch coverage, if the opposite of some branch, $b_i$, is not yet covered (and in the absence of any other information) then we should always try to flip $b_i$ because if successful, we are sure to cover an uncovered branch. However, if the opposite branch has already been covered then we must decide whether to flip $b_i$ in the hope that the suffix of the path covered by the newly generated case will contain some uncovered branch. We will call this *hopeful flipping*.

## D. Research question

This article investigates how to reduce hopeful flipping and tries to answer the following research question: ***Does reducing hopeful flipping in concolic generation for exhaustive branch coverage result in fewer solver calls?***

## II. OUR EXAMPLES

The strategies described here were tried on 7 real-life examples of C functions: A (a string-processing function from the Apache code), ANU (the same example with no unreachable branches), D (another string-processing function), E (the GNU Core Utils Echo utility), L (checks a property of credit-card numbers), T (the Tcas control logic) and TNU (the same example with no unreachable branches). The table shows the number of lines of code, branches, unreachable branches (with the given precondition) and loops (with a variable number of iterations) for each example.

|     | LOC | Branches | Unreachable | Loops |
| --- | --- | --- | --- | --- |
| A   | 70  | 30  | 1   | 4   |
| DNU | 70  | 20  | 0   | 1   |
| E   | 340 | 128 | 37  | 6   |
| L   | 50  | 18  | 2   | 2   |
| T   | 170 | 80  | 1   | 0   |

All the strategies except MT were run 10 times on each example and the results averaged, in an attempt to smooth the effects of non-determinism in constraint solving. The MT strategy, which combines the non-determinism of multi-threading with that of constraint solving, was run 100 times on each example.

## III. THE DFS STRATEGY

The first test generation strategy tried on our example is simple depth-first search of the FEP tree with a stopping criterion which is coverage of all the reachable branches.

As explained in the Introduction, on order to construct *pred(flip(p,i))*, the concolic test generator must traverse *pred(p)*, adding each constraint $c_0, c_1,...$ to the constraint satisfaction problem (CSP) until it reaches $c_i$, where it is $-c_i$, which is added. However, each newly covered path (except for the first) shares a prefix with the previously generated path. To avoid repeating the addition of the constraints of shared prefixes, we use incremental constraint solving and the state of the solver is stored in a stack after addition of each constraint. When a new test case is generated, covering a new path suffix, $s$, then instead of resubmitting the constraints from the shared prefix, we can just recover the state of the solver from the stack and start to add the constraints from $s$. In fact, for each $c_i$, we use backtracking to the previous state of the solver to alternate between flipping or not. If we choose to flip, then we add $-c_i$ to the CSP and then try to resolve it. If we choose not to flip then we add $c_i$ and proceed to the next constraint, $c_{i+1}$.

This is why it is particularly efficient to use an incremental solver and backtracking. To facilitate this, PathCrawler is implemented in Constraint Logic Programming (CLP) and uses the COLIBRI solver [7]. Note that adding a constraint to our CSP triggers *constraint propagation*. This has quadratic complexity but it enables certain forms of unsatisfiability to be detected before full constraint resolution, whose complexity is even greater. Below, we do not count constraint propagation as a solver call.

In our DFS strategy, the first branch to be flipped in each newly-covered path suffix, $s$, is the final branch. If this causes a new test case to be generated then the new suffix, $s'$, is then treated, as well as all suffixes generated from $s'$. After that, the generator backtracks back down $s$, flipping each branch in the same way. Test generation stops when either all branches have been covered or no branches are left to flip.

## IV. THE EAGER STRATEGY

The next strategy "reverses" DFS by systematically flipping each branch of $s$ first, and, if successful, exploring the new suffix, $s'$, before backtracking over the flip in order to proceed to the next branch in $s$. This makes the strategy more breadth-first and as a result, changes the order in which certain suffixes might be covered.

## V. THE LOOKAHEAD STRATEGY

This strategy tries to improve on Eager by taking account of whether flipping each branch could increase coverage. It is the same as Eager except that each branch is only flipped if its opposite is either uncovered or else could possibly lead to an uncovered branch. Whether the opposite branch could possibly lead to an uncovered branch is decided by a simple test of connectivity in the control-flow graph, without trying to evaluate the feasibility of the path to the uncovered branch.

## VI. THE ELSE STRATEGY

This strategy tries to improve on Lookahead by prioritising flipping of uncovered branches over hopeful flipping. It starts by eagerly flipping just the branches in $s$ whose opposite is uncovered and then backtracks down $s$ to hopefully flip any unflipped branches whose opposite may lead to uncovered branches.

## VII. THE MT STRATEGY

The two previous strategies modified Eager, which introduced an element of breadth-first search. However, Eager is not true breath-first search because the earlier branches in the suffix are still flipped before the later ones and suffixes covered by newly generated tests are explored before completing exploration of the suffixes covered by the earlier tests. In order to implement "fairer" breadth-first search strategies while still taking advantage of the efficiency of backtracking, we implemented Else using multi-threading.

In this strategy, each thread treats one covered path and each thread is treated in turn. Moreover, threads are classified as high-priority while the treatment advances along $s$ and then become low-priority when the treatment backtracks back down $s$ for hopeful flipping. Low-priority threads are only active while no threads are still high-priority. The information on which branches are covered is shared between threads.

## VIII. RESULTS

Exhaustive branch coverage was achieved on all examples and strategies. Figures 1 and 2 illustrate, on the different examples, the variation of hopeful flips and solver calls over the successive strategies. The value for the DFS strategy is taken as a reference and the values for the other strategies calculated as a percentage of this. Note that these DFS base values were much higher for A (1427 and 1719) than for ANU (152 and 165) but only slightly higher for T than for TNU.

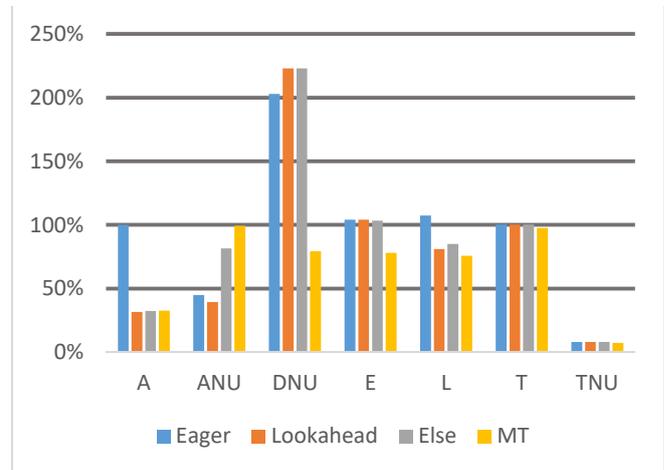

Fig. 1 Hopeful flips

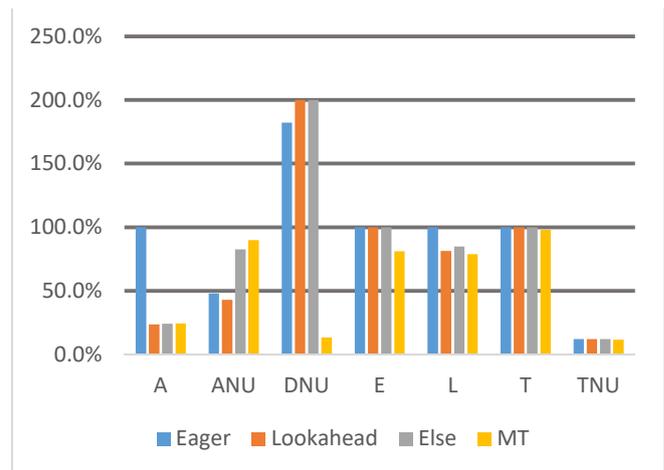

Fig. 2 Solver calls

The answer to our research question is that there is a correlation between hopeful flips and solver calls for all examples and strategies.

However, the influence of fortuitous coverage, and the way it varies on different examples, is also evident. It is the reason for which the results for Eager are sometimes worse than those for DFS and similarly for Else and Lookahead.

## IX. RELATED WORK

The work described here is related to a large body of work on devising symbolic-execution-based test generation strategies to limit *path explosion* [8]. Some researchers have proposed to tackle this problem with strategies to decide which branch to hopefully flip next, often based on heuristics, such as the shortest theoretical path to an uncovered branch [3][9]. Other approaches are based on a decomposition of the search space, for example by treating called functions separately and storing the results [10], or by identifying non-interfering blocks of code [11]. There has also been work on pruning the search space, by taking advantage of redundancies in the FEP tree to use information learnt during generation of previous tests [12][13][14]. Our work is complementary to most of this. It does propose basic pruning of the search space (based on connectivity in the control-flow graph) but tries above all to flip uncovered branches first, in order to limit the number of hopeful flipping choices, rather than focusing on how to choose which branch to hopefully flip next.

Our MT strategy is closely related to the forking of symbolic processes at the heart of the KLEE tool, which is not based on concolic generation. Instead, it builds a tree of symbolic processes which mirrors the FEP tree structure. At each node in the FEP tree, the feasibility of the prefix ending in one of the branches is checked and then, if feasible, that of the prefix ending in the other branch. If both are feasible, KLEE clones the symbolic execution state so that it can explore both paths. It is only at the end of a path, or when a possible assertion violation or run-time error is encountered, that it generates a test. KLEE has to make more feasibility checks (up to 2 per node of the FEP tree) than the concolic method and uses a constraint cache to limit solver calls. It does not benefit from incremental constraint solving or backtracking and this may make it less efficient [15] but breadth-first generation strategies are easily implemented. To do that in PathCrawler, while keeping the efficiency of backtracking, we used multi-threading. Our MT strategy clones not only the symbolic execution state, as in KLEE, but also the state of our incremental solver.

PathCrawler is implemented in CLP, which was also used to compare different test generation strategies to explore executable behavior models in [16]. They discuss the difficulty of implementing breadth-first strategies in CLP and instead of multi-threading, propose an interleaving strategy.

Unlike most previous work, ours explicitly takes account of how unreachable branches should be treated in exhaustive branch coverage.

## X. CONCLUSIONS

The answer to our Research Question - whether we can reduce the number of solver calls by reducing hopeful flipping in concolic test generation - was positive for our examples.

Our MT strategy is breadth-first, prioritises the flipping of uncovered branches and conditions the flipping of the other branches to control-flow graph connectivity to an uncovered branch. MT was designed to limit the number of hopeful flips when performing exhaustive branch coverage and successfully did so on all but the A and ANU examples. Indeeed, our experiments show that the effectiveness of concolic test generation strategies varies somewhat according to the tested function.

The next step in reducing the number of hopeful flips is by supplementary measures to prune the search space and this is what we will investigate in future work.